# Bayesian Method of Moments (BMOM) Analysis of Mean and Regression Models

Arnold Zellner[*]
University of Chicago

## Abstract

A Bayesian method of moments/instrumental variable (BMOM/IV) approach is developed and applied in the analysis of the important mean and multiple regression models. Given a single set of data, it is shown how to obtain posterior and predictive moments without the use of likelihood functions, prior densities and Bayes' Theorem. The posterior and predictive moments, based on a few relatively weak assumptions, are then used to obtain maximum entropy densities for parameters, realized error terms and future values of variables. Posterior means for parameters and realized error terms are shown to be equal to certain well known estimates and rationalized in terms of quadratic loss functions. Conditional maxent posterior densities for means and regression coefficients given scale parameters are in the normal form while scale parameters' maxent densities are in the exponential form. Marginal densities for individual regression coefficients, realized error terms and future values are in the Laplace or double-exponential form with heavier tails than normal densities with the same means and variances. It is concluded that these results will be very useful, particularly when there is difficulty in formulating appropriate likelihood functions and prior densities needed in traditional maximum likelihood and Bayesian approaches.

## 1 INTRODUCTION

In the traditional likelihood and Bayesian approaches, it is usually assumed that enough information is available to formulate a likelihood function and, in the Bayesian approach, a prior density for the parameters of the selected likelihood function—see e.g. Jeffreys (1988),

[*]Research financed in part by the National Science Foundation and by income from the H.G.B. Alexander Endowment Fund, Graduate School of Business, University of Chicago. Some of this work was reported to the U. of Chicago Econometrics and Statistics Colloquium in November, 1993, the Conference on Informational Aspects of Bayesian Statistics in Honor of H. Akaike, Japan, December 1993, the Department of Statistics Seminar, Iowa State U., April 1994, and the 1994 Conference in Honor of Seymour Geisser, Taiwan, December 1994. Thanks to participants and to a referee for helpful comments.

Box and Tiao (1973), Berger (1985), Geisser (1993), Press (1989), and Zellner (1971). However, if not enough information is available to specify a form for the likelihood function, then clearly there will be problems in both the traditional likelihood and Bayesian approaches. In situations like this, some resort to non-likelihood based methods, say least squares regression, method of moments or boot-strap approaches that are "data-based." That is, least squares is usually justified by its producing the best fit to a given sample of data with no appeal to sampling properties. However, if appropriate sampling assumptions are made, unbiasedness and a variance-covariance matrix of the least squares estimator can be produced without specifying a complete likelihood function. Similarly, the well-known Gauss-Markov theorem's assumptions provide certain optimality properties for the least square estimator without introducing a likelihood function. These general results, obtained under certain sampling assumptions have been widely utilized. However, they do not yield conditional probability statements about possible values of parameters based on a *single set of given observations*. In the present paper, it will be shown how such conditional results can be obtained without introducing sampling assumptions, likelihood functions and prior densities. Moments of parameters and future values of variables will be derived and shown to have certain optimal properties based on just a single set of observed data.

Further, to obtain a posterior distribution for parameters given just their moments, a maximum entropy (maxent) approach, see e.g. Jaynes (1982a,b) and Cover and Thomas (1991) will be utilized since this provides a most conservative choice of density that incorporates information in moment side conditions. However, if enough information is available to formulate a tentative likelihood function and a prior density for its parameters, the results of a traditional Bayesian analysis can be compared and/or combined with those yielded by the BMOM approach. See Green and Strawderman (1994) for an application of the BMOM approach in the analysis of a natural resource model with an unknown likelihood function.

The plan of the paper is as follows. In Section 2, an analysis of a simple scalar mean process is presented since this is a central, important case and analysis of it reveals well the essential features of the BMOM approach. Then in Section 3, results for multiple regression and autoregressive models, are given. Section 4 includes a summary of results and indications of future research.

## 2  ANALYSIS OF A SCALAR MEAN PROBLEM

In this section, we assume that n given observations $\underline{y}' = (y_1, y_2, .., y_n)$ have been obtained that relate to a scalar mean, $\theta$, as follows:

$$y_i = \theta + u_i \qquad i = 1, 2, ..., n \qquad (2.1)$$

where the $u_i$'s are unobserved, ***realized*** error terms; see Chaloner and Brant (1988), Zellner (1975) and Zellner and Moulton (1985) for traditional Bayesian analyses of realized error terms. Note that we have made the important assumption that the errors, say measurement errors, are additive. Since $\theta$ and the $u_i$'s are unobserved quantities with unknown values, we shall assume that we can view their possible values probabilistically. That is, given the data $\underline{y}$, we assume that the means of $\theta$ and the $u_i$'s as well as other features of their distributions exist but have unknown values. Note that a ***realized*** error term, $u_i$, usually does not have a zero mean. Operations and assumptions introduced below will enable us to express these posterior moments and other quantities as functions of the given data.

## 2.1 First Order Posterior and Predictive Moments

Given the observations' values in (2.1) both $\theta$ and the realized error terms, the $u_i$'s have fixed unknown values that we shall regard as subjectively random just as was done in Chaloner and Brant (1988), Zellner (1975), and Zellner and Moulton (1985). Here, in contrast to the work in these papers, we shall assume that not enough information is available to formulate a likelihood function and a prior density and thus it is not possible to use Bayes' theorem. From (2.1) we have

$$\bar{y} = \theta + \bar{u} \qquad (2.2)$$

where $\bar{y} = \sum_{i=1}^{n} y_i/n$, a given sample mean and $\bar{u} = \sum_{i=1}^{n} u_i/n$, the mean of the realized error terms. The symbol E denotes a posterior expectation operator, so-called because it is utilized after the data have been observed. From (2.2), we have

$$\bar{y} = E\theta|D + E\bar{u}|D \qquad (2.3)$$

where D denotes the given data, here $(y_1, y_2, ..., y_n)$.

We now introduce the following assumption.

*Assumption I*: $\qquad E\bar{u}|D = \sum_{i=1}^{n} Eu_i|D/n = 0 \qquad (2.4)$

This assumption indicates that we believe that there is nothing systematic in the realized error terms and thus the expectation of their mean is equal to zero. If, for example, we believe that the i'th observation is an additive outlier, we would have $u_i = \eta + \varepsilon_i$ with the mean of the parameter $\eta$, $E\eta|D \neq 0$ and then $E\bar{u}|D \neq 0$.

Given Assumption I in (2.4), we have from (2.3),

$$E\theta|D = \bar{y} \qquad (2.5)$$

that is the posterior mean of $\theta$ is the sample mean $\bar{y}$. Note that this result has been obtained without selecting a likelihood function and a prior density for its parameters as is done in many analyses.

Given the result in (2.5), we have on taking the expectation of both sides of (2.1), the posterior expectation of $u_i$, the i'th realized error term is:

$$Eu_i|D = y_i - E\theta|D = y_i - \bar{y} \qquad (2.6)$$

which is the deviation of $y_i$ from the mean $\bar{y}$. On summing both sides of (2.6) and dividing by n, we have $\sum_{i=1}^{n} Eu_i|D/n = \sum_{i=1}^{n} (y_i - \bar{y})/n = 0$, the "sample analogue" of Assumption I.

Further, note that if we seek an optimal point estimate, $\tilde{\theta}$, relative to a quadratic loss function $L(\theta,\tilde{\theta}) = (\theta-\tilde{\theta})^2$, we have $EL(\theta,\tilde{\theta}) = E(\theta-E\theta)^2 + (\tilde{\theta}-E\theta)^2$ and as is well known, taking $\tilde{\theta} = E\theta$ is optimal. Thus the estimate, $\bar{y}$ in (2.5) is optimal relative to squared error loss.

If $y_{n+1}$ is a future, as yet unobserved value satisfying $y_{n+1} = \theta + u_{n+1}$, where $u_{n+1}$ is

-3-

a future, as yet unrealized error term, we can write $Ey_{n+1}|D = E\theta|D + Eu_{n+1}|D$ and assume that $Eu_{n+1}|D = 0$. Then

$$Ey_{n+1}|D = E\theta|D = \bar{y} \tag{2.7}$$

which is the mean of the future observation $y_{n+1}$ given the data $\underline{y}$ and the information in Assumption I. $\bar{y}$ is an optimal point prediction for $y_{n+1}$ relative to a squared error predictive loss function.

The first order moments in (2.5), (2.6) and (2.7) are identical to those obtained in a traditional normal likelihood function, diffuse prior analysis. However (2.5), (2.6) and (2.7) have been obtained without an iid normality assumption and without an improper diffuse prior.

## 2.2 Second Order Posterior and Predictive Moments

Having derived the posterior and predictive means in (2.5) and (2.7), we now turn to consider derivation of second order posterior and predictive moments.

From (2.1) in vector notation $\underline{y} = \underline{\iota}\theta + \underline{u}$ where $\underline{\iota}' = (1, 1, ..., 1)$ a $1 \times n$ vector of ones, we have $\hat{\underline{u}} = \underline{y} - \underline{\iota}\bar{y} = [I - \underline{\iota}(\underline{\iota}'\underline{\iota})^{-1}\underline{\iota}']\underline{y} = [I - \underline{\iota}(\underline{\iota}'\underline{\iota})^{-1}\underline{\iota}']\underline{u}$ and thus

$$\underline{u} - \hat{\underline{u}} = \underline{\iota}(\underline{\iota}'\underline{\iota})^{-1}\underline{\iota}'\underline{u} = \underline{\iota}(\underline{\iota}'\underline{\iota})^{-1}\underline{\iota}'(\underline{u}-\hat{\underline{u}}) \tag{2.8}$$

where $\underline{\iota}'\hat{\underline{u}} = 0$ has been employed in (2.8). Then,

$$V(\underline{u}|D) = E(\underline{u}-\hat{\underline{u}})(\underline{u}-\hat{\underline{u}})'|D = \underline{\iota}(\underline{\iota}'\underline{\iota})^{-1}\underline{\iota}'E(\underline{u}-\hat{\underline{u}})(\underline{u}-\hat{\underline{u}})'|D\underline{\iota}(\underline{\iota}'\underline{\iota})^{-1}\underline{\iota}' \tag{2.9}$$

is a functional equation that $V(\underline{u}|D)$ must satisfy, where D denotes given sample and prior information. Thus we introduce the following assumption.

*Assumption II*: $\quad V(\underline{u}|\sigma^2,D) = E(\underline{u}-\hat{\underline{u}})(\underline{u}-\hat{\underline{u}})'|\sigma^2,D = \underline{\iota}(\underline{\iota}'\underline{\iota})^{-1}\underline{\iota}'\sigma^2 \tag{2.10}$

with $\sigma^2$ a positive scalar parameter which satisfies (2.9). Note from $y_i = \theta + u_i$ and $y_i = \bar{y} + \hat{u}_i$ that $u_i - \hat{u}_i = \bar{y}-\theta$. Also, $u_j-\hat{u}_j = \bar{y}-\theta$ and thus $u_i-\hat{u}_i = u_j-\hat{u}_j$ for all i,j. Thus it is not surprising that in (2.10) all variances are the same and all correlations equal 1. Then from $\bar{y} - \theta = (\underline{\iota}'\underline{\iota})^{-1}\underline{\iota}'(\underline{y}-\underline{\iota}\theta) = (\underline{\iota}'\underline{\iota})^{-1}\underline{\iota}'(\underline{u}-\hat{\underline{u}})$, we have

$$V(\theta|\sigma^2,D) = E(\theta-\bar{y})^2|\sigma^2,D = (\underline{\iota}'\underline{\iota})^{-1}\underline{\iota}'E(\underline{u}-\hat{\underline{u}})(\underline{u}-\hat{\underline{u}})'\underline{\iota}(\underline{\iota}'\underline{\iota})^{-1}|\sigma^2,D \tag{2.11}$$
$$= (\underline{\iota}'\underline{\iota})^{-1}\sigma^2 = \sigma^2/n$$

where Assumption II has been used for $E(\underline{u}-\hat{\underline{u}})(\underline{u}-\hat{\underline{u}})'|\sigma^2,D$ in the first line of (2.11). Thus $\sigma^2/n$ is the posterior variance for $\theta$ given the parameter $\sigma^2$ and the data.

In addition, $E(\underline{u}-\hat{\underline{u}})'(\underline{u}-\hat{\underline{u}})|D,\sigma^2 = \text{tr } \underline{\iota}(\underline{\iota}'\underline{\iota})^{-1}\underline{\iota}'\sigma^2 = \sigma^2$ from (2.10). Then $E(\underline{u}-\hat{\underline{u}})'(\underline{u}-\hat{\underline{u}})|D/n = E\underline{u}'\underline{u}/n|D - \hat{\underline{u}}'\hat{\underline{u}}/n = E\sigma^2|D/n$ where $E\underline{u}|D = \hat{\underline{u}}$ has been used. If we define $E\sigma^2|D = E\sum_{j=1}^{n}(u_j - E\bar{u})^2/n|D = E\underline{u}'\underline{u}/n|D$, since $E\bar{u}|D = 0$ by Assumption I, we have $E\sigma^2|D - \hat{\underline{u}}'\hat{\underline{u}}/n = E\sigma^2/n|D$ or

$$E\sigma^2|D = \hat{\underline{u}}'\hat{\underline{u}}/(n-1) \equiv s^2 \tag{2.12}$$

-4-

which is the posterior mean of $\sigma^2$ and thus $V(\theta|D) = s^2/n$.

For the future observation $y_{n+1} = \theta + u_{n+1}$, $Ey_{n+1} = \bar{y}$, given $E\theta = \bar{y}$ and $Eu_{n+1}|D,\sigma^2 = 0$. Also, $y_{n+1} - \bar{y} = \theta - \bar{y} + u_{n+1}$ and

$$E[(y_{n+1}-\bar{y})^2|\sigma^2,D] = E(\theta-\bar{y})^2|\sigma^2,D + Eu_{n+1}^2|\sigma^2,D = \sigma^2(1+1/n) \qquad (2.13)$$

given that $u_{n+1}$ and $\theta-\bar{y}$ are uncorrelated, (2.11) and $Eu_{n+1}^2|D,\sigma^2 = \sigma^2$.

A traditional diffuse prior, normal likelihood approach produces results identical to those in (2.11) and (2.13). However, the traditional approach yields $E(\sigma^2|D) = \nu s^2/(\nu-2)$ for $\nu > 2$ where $\nu = n-1$ rather than (2.12), $E(\sigma^2|D) = s^2$. Also $V(\theta|D) = s^2/n$ in the BMOM approach, whereas in the traditional Bayesian approach $V(\theta|D) = \nu s^2/(\nu-2)n$. Further, (2.1), (2.12) and (2.13) are obtained without an iid normality assumption and an improper prior density.

## 2.3 Derivation of Posterior and Predictive Densities

Above we have derived the first two posterior moments of $\theta$ given $\sigma^2$, namely $E\theta|D = \bar{y}$ and $Var(\theta|\sigma^2,D) = \sigma^2/n$. Also, $E\sigma^2|D = s^2 = \underline{\hat{u}}'\underline{\hat{u}}/(n-1)$, $V(\theta|D) = s^2/n$, $Ey_{n+1}|D = \bar{y}$ and $Var(y_{n+1}|\sigma^2,D) = (1+1/n)\sigma^2$. It is well known that maxent densities can be derived that incorporate the information in moment conditions; see e.g. Jaynes (1982a,b), Cover and Thomas (1991) and Zellner (1991, 1993). That is, we choose a density, say f(x), to maximize $H(f) = -\int f(x)\ell nf(x)dx$ subject to $\int x^i f(x)dx = \mu_i$, $i = 0, 1, 2, ....$ with $\mu_0 = 1$, where we have utilized uniform measure in defining entropy, H(f). See Shore and Johnson (1980) for consistency, invariance and other desirable properties of this entropy-based procedure. Here we have the following results.

A. The proper maxent posterior density for $\theta$ given $\sigma^2$ is a normal density with mean $\bar{y}$ and variance $\sigma^2/n$, i.e. $g_N(\theta|\sigma^2,D) \sim N(\bar{y},\sigma^2/n)$.

B. The proper maxent predictive density for $y_{n+1}$ given $\sigma^2$ and D with mean $\bar{y}$ and variance $\sigma^2(1+1/n)$ is a normal density $h_N(y_{n+1}|\sigma^2,D)$ with these moments, $N[\bar{y},\sigma^2(1+1/n)]$.

C. The proper maxent posterior density for $\sigma^2$ with $E\sigma^2 = s^2$ is the exponential density $g_e(\sigma^2|D) = (1/s^2)\exp\{-\sigma^2/s^2\}$, $0 < \sigma^2 < \infty$, with $s^2$ given in (2.12).

D. From A and C, the joint posterior density for $\theta$ and $\sigma^2$ is

$$f(\theta,\sigma^2|D) = g_N(\theta|\sigma^2,D)g_e(\sigma^2|D) \qquad (2.14)$$

which is a maxent density given that $\theta/\sigma$ and $\sigma$ are assumed independent.[1] On integrating over $\sigma^2$, $0 < \sigma^2 < \infty$, the marginal posterior density for $\theta$ is—see Appendix for the derivation,

$$g(\theta|D) = \sqrt{n/2s^2} \exp\left\{-\sqrt{2n}\,|\theta-\bar{y}|/s\right\} \qquad -\infty < \theta < \infty \quad (2.15)$$

This double-exponential or Laplace[2] marginal posterior density for $\theta$ is symmetric about $\bar{y}$, the mean, median and modal value, with variance equal to $s^2/n$ and has thick tails relative

---

[1]A maxent density without this independence assumption has also been derived.

[2]See Stigler (1986, p.111) for a fascinating discussion of Laplace's derivation of this distribution using the "principle of insufficient reason."

-5-

to normal or Student-t densities. By a change of variable, $z = \sqrt{n}\,(\theta-\bar{y})/s$, a standardized form of (2.15) is: $g(z|D) = (1/\sqrt{2})\exp\{-\sqrt{2}\,|z|\}$, $-\infty < z < \infty$. See Appendix for further properties of this density.

E. From B and C, the joint density for $y_{n+1}$ and $\sigma^2$ is

$$h_N(y_{n+1}|\sigma^2,D)g_e(\sigma^2|D) \tag{2.16}$$

where $h_N \sim N[\bar{y}, \sigma^2(1+1/n)]$ and $g_e$ is the exponential density shown in C. On integrating (2.16) with respect to $\sigma^2$, $0 < \sigma^2 < \infty$, the marginal predictive density of $y_{n+1}$ is in the following double-exponential form,

$$h_e(y_{n+1}|D) = \left(1/s_e\sqrt{2}\right)\exp\left\{-\sqrt{2}\,|y_{n+1}-\bar{y}|/s_e\right\} \qquad -\infty < y_{n+1} < \infty \tag{2.17}$$

where $s_e^2 = (1+1/n)s^2$. The mean, median and modal value of (2.17) are all equal to $\bar{y}$ and its variance is $s_e^2$. As with the density in (2.15), the tails of the predictive density can be rather thick.

F. The proper maxent posterior density for $u_i = y_i - \theta$, the i'th realized error term with mean $\hat{u}_i$ and variance $\sigma^2/n$ is given by

$$g_N(u_i|\sigma,D) = \sqrt{n/2\pi\sigma^2}\,\exp\left\{-(u_i-\hat{u}_i)^2 n/2\sigma^2\right\} \qquad -\infty < u_i < \infty \tag{2.18}$$

G. The marginal posterior density for $u_i$, obtained by integration from the joint density, $g_N(u_i|\sigma,D)g_e(\sigma^2|D)$ is

$$g(u_i|D) = \sqrt{n/2s^2}\,\exp\left\{-\sqrt{2n}\,|u_i-\hat{u}_i|/s\right\} \qquad -\infty < u_i < \infty \tag{2.19}$$

a double-exponential density centered at $\hat{u}_i$, the i'th residual.

H. If it is known that $y_i > 0$ for all i, as with "time to failure" or "duration" data, $0 < \theta < \infty$ and the maxent proper density for $\theta$ subject just to $E\theta = \bar{y} > 0$ is the following exponential density

$$g_1(\theta|D) = (1/\bar{y})\exp\{-\theta/\bar{y}\} \qquad 0 < \theta < \infty \tag{2.20}$$

The posterior and predictive densities above can readily be implemented in practice. As indicated below, they can be compared and/or combined with posterior and predictive densities derived from assumed likelihood functions and prior densities using Bayes' theorem.

# 3 BMOM ANALYSIS OF MULTIPLE REGRESSION AND AUTO-REGRESSION MODELS

The BMOM analysis of multiple regression and autoregression models is quite similar to that utilized in analyzing the scalar mean model in the previous section. The given $n \times 1$ observation vector $\underline{y}$ is assumed to satisfy the following n well known equations,

$$\underline{y} = X\underline{\beta} + \underline{u} \tag{3.1}$$

where X is a given $n \times k$ matrix of rank k, $\underline{\beta}$ is a $k \times 1$ vector of regression coefficients with unknown values and $\underline{u}$ is an n × 1 vector of realized error terms. If (3.1) relates to an autoregression, it is assumed that the initial values of the output variable, say $(y_0, y_{-1}, ..., y_{-(q-1)})$, for a q'th order autoregression are known and are elements of the X matrix. Given that it is assumed that we do not have enough information to formulate a likelihood function with much confidence, the BMOM approach can be employed to derive posterior and predictive moments and densities.

## 3.1 Derivation of First Order Posterior and Predictive Moments

To derive first order moments, we multiply both sides of (3.1) by $(X'X)^{-1}X'$ and take the posterior expectation of both sides to obtain, as in connection with (2.2),

$$(X'X)^{-1}X'\underline{y} = E\underline{\beta}|D + (X'X)^{-1}X'E\underline{u}|D \qquad (3.2)$$

Now paralleling Assumption I above in (2.4), the following Assumption I' is introduced.

*Assumption I':*  $\qquad\qquad (X'X)^{-1}X'E\underline{u}|D = \underline{0}$ $\qquad\qquad (3.3)$

If one of the columns of X is a column of ones, that is there is an intercept in (3.10), (3.3) includes Assumption I, namely $E\bar{u}|D = \sum_1^n Eu_i|D/n = 0$. Further, (3.3) implies that there is nothing "systematic" in $E\underline{u}|D$ that is correlated with the columns of X. For example, this would not be the case if it were believed that some important variable or variables had been omitted from the relation in (3.1) or if it were believed that the values of the independent variables in X were measured with error. In these two cases, the $u_i$'s, would contain components that would be expected to be correlated with columns of X and (3.3) would not be expected to hold. However, if it is believed that (3.3) is valid, then clearly from (3.2) the posterior mean, $E\underline{\beta}$ is given by

$$E\underline{\beta}|D = (X'X)^{-1}X'\underline{y} = \underline{\hat{\beta}} \qquad (3.4)$$

which is the least squares quantity, a rather simple result[3]. Further, given a quadratic loss function, $L(\underline{\beta},\underline{\tilde{\beta}}) = (\underline{\beta}-\underline{\tilde{\beta}})'Q(\underline{\beta}-\underline{\tilde{\beta}})$, where Q is a given positive definite symmetric matrix, the value of $\underline{\tilde{\beta}}$ that minimizes posterior expected loss is the posterior mean in general and given in (3.4) for this specific problem.

From (3.4), the mean of the realized error vector is given by

$$E\underline{u}|D = \underline{y} - XE\underline{\beta}|D = \underline{y} - X\underline{\hat{\beta}} = \underline{\hat{u}} \qquad (3.5)$$

where $\underline{\hat{u}}$ is the least squares residual vector and we have $X'\underline{\hat{u}} = 0$, the sample analogue of Assumption I' in (3.3).

As regards the predictive mean, we have for a future observation, $y_f$, assumed to satisfy $y_f = \underline{x}_f'\underline{\beta} + u_f$, with the 1 × k vector $\underline{x}_f'$ given and $u_f$ a random, as yet unrealized error term assumed drawn from a distribution with a zero mean (equal to $E\bar{u}|D = 0$, as assumed above in Assumption I'). Then the predictive mean is

$$Ey_f|D = x_f'E\beta|D = \underline{x}_f'\hat{\beta} = \hat{y}_f \tag{3.6}$$

which is just the least squares point prediction. Given a squared error predictive loss function, the predictive mean in (3.6) minimizes the expectation of such a predictive loss function.

---

[3]If, for example, rather than (3.1), we have $\underline{y} = X\underline{\beta} + Z\underline{\gamma} + \underline{\varepsilon}$, that is $\underline{u} = Z\underline{\gamma} + \underline{\varepsilon}$, $\underline{\tilde{\beta}} = (X'X)^{-1}X'\underline{y} = \underline{\beta} + (X'X)^{-1}X'Z\underline{\gamma} + (X'X)^{-1}X'\underline{\varepsilon}$ and $\underline{\tilde{\beta}} = E\underline{\beta}|D + (X'X)^{-1}X'ZE\underline{\gamma}|D$ given $(X'X)^{-1}X'E\underline{\varepsilon}|D = \underline{0}$. Thus $E\underline{\beta}|D \neq \underline{\tilde{\beta}}$ for $X'Z \neq 0$ and $E\underline{\gamma}|D \neq 0$.

Having obtained the first order moments above, we now turn to derive second order moments.

## 3.2 Derivation of Second Order Posterior and Predictive Moments

From (3.1) and (3.4), we have $\underline{\beta} - E\beta|D = (X'X)^{-1}X'(\underline{u}-\hat{u})$, where $\hat{u} = \underline{y}-X\hat{\beta}$ with $X'\hat{u} = 0$. Thus the posterior covariance matrix, denoted by $V(\beta|D)$, is

$$V(\beta|D) = E(\underline{\beta}-\hat{\beta})(\underline{\beta}-\hat{\beta})'|D = (X'X)^{-1}X'E(\underline{u}-\hat{u})(\underline{u}-\hat{u})'|DX(X'X)^{-1} \tag{3.7}$$

To evaluate (3.7) another assumption is needed, paralleling that in (2.10). Given that $\hat{u} = (I - X(X'X)^{-1}X')\underline{y} = (I - X(X'X)^{-1}X')\underline{u}$, we have that

$$\underline{u} - \hat{u} = X(X'X)^{-1}X'(\underline{u}-\hat{u}) \tag{3.8}$$

and

$$E(\underline{u}-\hat{u})(\underline{u}-\hat{u})'|D = X(X'X)^{-1}X'E(\underline{u}-\hat{u})(\underline{u}-\hat{u})'|DX(X'X)^{-1}X' \tag{3.9}$$

is a functional equation that must be satisfied. Note that only k elements of $\underline{u}$ are free. Thus, we introduce the following assumption, the analogue of (2.10):

*Assumption II':* $\quad V(\underline{u}|\sigma^2,D) = E(\underline{u}-\hat{u})(\underline{u}-\hat{u})'|\sigma^2,D = X(X'X)^{-1}X'\sigma^2 \tag{3.10}$

where $\sigma^2$ is a positive constant.[4] On inserting the expression in (3.10) in (3.9), it is seen that the functional equation is satisfied for any given value of $\sigma^2$. Substituting from (3.10) in (3.7), we have

$$V(\beta|\sigma^2,D) = (X'X)^{-1}\sigma^2 \tag{3.11}$$

which is the posterior covariance matrix for $\underline{\beta}$ given $\sigma^2$ and D, the data.

To obtain a value for the posterior expectation of $\sigma^2$, from (3.10), $E(\underline{u}-\hat{u})'(\underline{u}-\hat{u})|\sigma^2,D = \text{tr } X(X'X)^{-1}X'\sigma^2 = k\sigma^2$. Then $E\underline{u}'\underline{u}/n|D - \hat{u}'\hat{u}/n = kE\sigma^2|D/n$ and if we define $E\sigma^2|D = E\sum_{i=1}^n (u_i-E\bar{u})^2/n|D = E\underline{u}'\underline{u}/n|D$, since $E\bar{u}|D = 0$ from Assumption I', $E\sigma^2|D - \hat{u}'\hat{u}/n = kE\sigma^2|D/n$ or

$$E\sigma^2|D = \hat{u}'\hat{u}/(n-k) = s^2 \tag{3.12}$$

which is the posterior mean of $\sigma^2$.

For the future observation, $y_f = \underline{x}_f'\underline{\beta} + u_f$, $Ey_f|D = x_f'\hat{\beta} \equiv \hat{y}_f$, as shown above. Then $y_f - \hat{y}_f = \underline{x}_f'(\underline{\beta}-\hat{\beta}) + u_f$ and

$$E(y_f - \hat{y}_f)^2 | \sigma^2, D = (1 + \underline{x}_f'(X'X)^{-1}\underline{x}_f)\sigma^2 \qquad (3.13)$$

given that $u_f$ and the elements of $\underline{\beta} - \underline{\hat{\beta}}$ have zero covariances and $Eu_f | D = 0$.

---

[4]If we write $\underline{y} = XHH^{-1}\underline{\beta} + \underline{u} = Z\underline{\theta} + \underline{u}$, where H is a square k×k non-singular matrix such that $H'X'XH = I_k$ and $\underline{\theta} = H^{-1}\underline{\beta}$, a k×1 vector. Then $Z'\underline{y} = \underline{\theta} + Z'\underline{u}$ and $E\underline{\theta}|D = Z'\underline{y}$ given $Z'E\underline{u}|D = 0$. Also, if we assume that the k error terms in $Z'\underline{u}$ have equal variances and are mutually uncorrelated, this implies (3.10) and (3.11).

Given the above posterior and predictive moments, the following are maxent posterior and predictive densities that incorporate the information in these moments.

A'. The proper maxent posterior density for $\underline{\beta}$ given $\sigma^2$ and the data with mean $\underline{\hat{\beta}} = (X'X)^{-1}X'\underline{y}$ and covariance matrix $(X'X)^{-1}\sigma^2$, is a multivariate normal density, $g_N(\underline{\beta}|\sigma^2,D) \sim MVN(\underline{\hat{\beta}}, (X'X)^{-1}\sigma^2)$.

B'. The proper maxent predictive density for $y_f$ given $\sigma^2$, $\underline{x}_f$ and the data is a normal density, $h_N(y_f|\sigma^2,D) \sim N[(\hat{y}_f, (1 + \underline{x}_f'(X'X)^{-1}\underline{x}_f)\sigma^2)]$.

C'. The proper maxent density for $\sigma^2$ with $E\sigma^2 = s^2 = \underline{\hat{u}}'\underline{\hat{u}}/(n-k)$ is the exponential density $g_e(\sigma^2|D) = (1/s^2)\exp(-\sigma^2/s^2)$, $0 < \sigma^2 < \infty$.

D'. From A' and C', the joint posterior density for $\underline{\beta}$ and $\sigma^2$ is[5]

$$f(\underline{\beta},\sigma^2|D) = g_N(\underline{\beta}|\sigma^2,D)g_e(\sigma^2|D) \qquad (3.14)$$

which is a maxent density given that $\underline{\beta}/\sigma$ and $\sigma$ are independent. The marginal posterior density of a single element of $\underline{\beta}$, say $\beta_i$, can be obtained by integrating (3.14) with respect to the remaining elements of $\underline{\beta}$ and then with respect to $\sigma^2$. The result is the following double exponential density for $\beta_i$:

$$g(\beta_i|D) = \left(1/s_i\sqrt{2}\right)\exp\left\{-\sqrt{2}\,|\beta_i - \hat{\beta}_i|/s_i\right\} \qquad -\infty < \beta_i < \infty \qquad (3.15)$$

where $\hat{\beta}_i$ is the i'th element of $\underline{\hat{\beta}}$ and $s_i^2$ is the (i,i)'th element of $(X'X)^{-1}s^2$. Also, from (3.14), the marginal distribution of $\eta = \underline{\ell}'\underline{\beta}$, where $\underline{\ell}$ is a given vector of rank one, is

$$g(\eta|D) = \left(1/s_\eta\sqrt{2}\right)\exp\left\{-\sqrt{2}\,|\eta - \hat{\eta}|/s_\eta\right\} \qquad -\infty < \eta < \infty \qquad (3.16)$$

where $\hat{\eta} = \underline{\ell}'\underline{\hat{\beta}}$ and $s_\eta^2 = \underline{\ell}'(X'X)^{-1}\underline{\ell}\,s^2$

E'. From B' and C', the joint density for $y_f$ and $\sigma$ is

$$h_N(y_f|\sigma^2,D)g_e(\sigma^2|D) \qquad (3.17)$$

where $h_N \sim N[\hat{y}_f, (1 + \underline{x}_f'(X'X)^{-1}\underline{x}_f)\sigma^2]$ and $g_e$ is the exponential density shown in C'. On integrating (3.17) with respect to $\sigma^2$, the marginal predictive density for $y_f$ is the following double-exponential density:

$$h_f(y_f|D) = \left(1/s_e\sqrt{2}\right)\exp\left\{-\sqrt{2}\,|y_f - \hat{y}_f|/s_e\right\} \qquad -\infty < y_f < \infty \qquad (3.18)$$

The mean of this density is $\hat{y}_f$ and its variance is $s_e^2 = [1 + \underline{x}_f'(X'X)^{-1}\underline{x}_f]s^2$.

F'. The proper maxent posterior density for $u_i$ with given mean $\hat{u}_i$ and given variance $v_i = \underline{x}_i'(X'X)^{-1}\underline{x}_i\sigma^2$ is

$$g_N(u_i|\sigma^2,D) = 1/\sqrt{2\pi v_i}\,\exp\{-(u_i-\hat{u}_i)^2/2v_i\} \qquad (3.19)$$

---

[5]To compute the joint density of the elements of $\underline{\beta}$, draw $\sigma^2$ from $g_e(\sigma^2|D)$ and insert the drawn value in $g_N(\underline{\beta}|\sigma^2,D)$ and draw $\underline{\beta}$ from this multivariate normal density. Thus draws from the joint density $f(\underline{\beta},\sigma^2|D)$ are obtained by repeated use of this procedure.

G′. The marginal posterior density for $u_i$, obtained by integrating the joint density $g_N(u_i|\sigma,D)g_e(\sigma^2|D)$ with respect to $\sigma$ is:

$$g_e(u_i|D) = \left[1/\sqrt{2v_i}\right]\exp\left\{-\sqrt{2}\,|u_i-\hat{u}_i|/\sqrt{v_i}\right\} \qquad -\infty < u_i < \infty \qquad (3.20)$$

a double-exponential density.

H′. If it is known that $\theta = \underline{\ell}'\underline{\beta}$ is strictly positive, where $\underline{\ell}'$ is a given $1 \times k$ vector of rank one, the maxent density for $\theta$ subject just to $E\theta|D = \underline{\ell}'\underline{\hat{\beta}} > 0$, is the following exponential density:

$$g(\theta|D) = (1/\underline{\ell}'\underline{\hat{\beta}})\exp\{-\theta/\underline{\ell}'\underline{\hat{\beta}}\} \qquad 0 < \theta < \infty \qquad (3.21)$$

## 3.3 Use of Additional Prior Information

If in addition to the sample information $\underline{y} = X\underline{\beta} + \underline{u}$, we represent additional prior information by use of a conceptual sample, $\underline{y}_c = X_c\underline{\beta} + \underline{u}_c$, where $\underline{y}_c$ is an $n_c \times 1$ vector of conceptual data points, $X_c$ is a $n_c \times k$ given matrix, $\underline{\beta}$ is a $k \times 1$ vector of regression parameters and $\underline{u}_c$ is an $n_c \times 1$ vector of realized conceptual error terms. We can write

$$\begin{pmatrix}\underline{y}_c \\ \underline{y}\end{pmatrix} = \begin{pmatrix}X_c \\ X\end{pmatrix}\underline{\beta} - \begin{pmatrix}\underline{u}_c \\ \underline{u}\end{pmatrix} \qquad (3.22a)$$

or

$$\underline{w} = W\underline{\beta} + \underline{\varepsilon} \qquad (3.22b)$$

Then making Assumption I′ and II′ relative to the system in (3.22b), the posterior mean of $\underline{\beta}$ is

$$\begin{aligned}\underline{\tilde{\beta}} \equiv E\underline{\beta}|D' &= (W'W)^{-1}W'\underline{w} \qquad (3.23)\\ &= (X_c'X_c + X'X)^{-1}(X_c'\underline{y}_c + X'\underline{y})\\ &= (X_c'X_c + X'X)^{-1}(X_c'X_c\underline{\hat{\beta}}_c + X'X\underline{\hat{\beta}})\end{aligned}$$

where, when $X_c$ is of full column rank, $\underline{\hat{\beta}}_c = (X_c'X_c)^{-1}X_c'\underline{y}_c$ is a prior mean vector, $\underline{\hat{\beta}} = (X'X)^{-1}X'\underline{y}$, the least squares estimate and $X_c'X_c$ is assigned a value by the investigator. Then, with Assumption II′ relating to (3.22b), we have

$$E(\underline{\beta}-E\underline{\beta})(\underline{\beta}-E\underline{\beta})'|D,\sigma^2) = (W'W)^{-1}\sigma^2 \qquad (3.24)$$

and

$$E\sigma^2|D = (\underline{w}-W\underline{\tilde{\beta}})'(\underline{w}-W\underline{\tilde{\beta}})/(n+n_c-k). \qquad (3.25)$$

Also maxent distributions are available for this system that incorporates a conceptual sample.

Above are some results of applying the BMOM approach to analyze data assumed generated by a multiple regression or an autoregression model. In addition, posterior and predictive intervals can easily be computed from the above posterior and predictive densities by the procedures described in the Appendix. Generally these intervals are broader than corresponding intervals based on the conditional normal posterior and predictive densities, derived above, with $\sigma^2 = s^2$. Also they are broader than conventional intervals based on marginal Student-t densities based on posterior and predictive densities based on normal likelihood functions and diffuse prior densities for their parameters. See Appendix A for one such comparison.

# 4 Summary and Concluding Remarks

In the preceding sections, posterior and predictive moments for parameters and future, as yet unobserved observations have been derived using one given set of data and a few simple assumptions. There was no need to formulate a density function for as yet unobserved data as a basis for a likelihood function nor to introduce a prior density for its parameters. Also, no use was made of Bayes' Theorem in deriving posterior and predictive moments.

Then, proper posterior and predictive densities were derived by maximizing entropy subject to the derived moments. These densities for location or regression coefficients with doubly infinite ranges are in the double exponential or Laplacian form while the maxent posterior density for variance parameters are in the exponential form. For location parameters with strictly positive values, the maxent posterior densities subject just to a first moment constraint are in the exponential form. Similar results were obtained for predictive densities for as yet unobserved observations.

It should be appreciated that these maxent predictive densities, in the double exponential or exponential form can serve as models for as yet unobserved data and employed in calculation of posterior odds in model comparison and selection problems using new data. In particular such models can be compared to those derived using a particular likelihood function, a prior density for its parameters and Bayes' Theorem. The two different posterior densities, based on analyses of a given sample of data can be employed as prior densities in the calculation of posterior odds using a second sample of data. The posterior odds can then be used to choose between the two models or to combine them using the approach described in Min and Zellner (1993) and Palm and Zellner (1992). In future work, such calculations will be reported. See Zellner (1995) for some BMOM results relating to multivariate regression and simultaneous equations models.

# Appendix
## Derivation of Double Exponential Density in (2.15)

From (2.14), $f(\theta,\sigma^2|D) = g_N(\theta|\sigma^2,D)g_e(\sigma^2|D)$ with $g_N(\theta|\sigma^2,D) = \sqrt{n/2\pi\sigma^2} \times \exp\{-n(\theta-\bar{y})^2/2\sigma^2\}$ and $g_e(\sigma^2|D) = (1/s^2)\exp\{-\sigma^2/s^2\}$. Then, with $a \equiv 1/s^2$ and $b \equiv n(\theta-\bar{y})^2/2$

$$\int_0^\infty f(\theta,\sigma^2|D)d\sigma^2 = \sqrt{n/2\pi}\, a \int_0^\infty \frac{1}{\sigma}\exp\{-[b/\sigma^2 + a\sigma^2]\}d\sigma^2$$

$$= \sqrt{2n/\pi}\, a \int_0^\infty \exp\{-[b/\sigma^2 + a\sigma^2]\}d\sigma \quad\quad (A.1)$$

$$= \sqrt{2n/\pi}\, a \left[\frac{1}{2}\sqrt{\pi/a}\exp\{-2\sqrt{ab}\}\right]$$

where the integral has been evaluated using a result given in Gradshteyn and Ryzhik (1980, p. 307, entry 3.325). On inserting the above values for a and b in (A.1), we have

$$g(\theta|D) = \sqrt{n/2s^2}\, \exp\{-2\sqrt{n/2}\,|\theta-\bar{y}|/s\} \qquad -\infty < \theta < \infty \qquad (A.2)$$

Further, letting $w = \sqrt{n/2}(\theta-\bar{y})/s$, a "standardized" form of the density is

$$g(w|D) = \exp\{-2|w|\} \qquad -\infty < w < \infty \qquad (A.3)$$

Note that $\int_0^\infty \exp\{-2w\}dw = \left[-\tfrac{1}{2}\exp(-2w)\right]_0^\infty = 1/2$ and, given symmetry of $g(w|D)$, $\int_{-\infty}^\infty g(w|D)dw = 1$. Also, the symmetry of $g(w|D)$ implies that all odd order moments about zero are zero; that is, $\int_{-\infty}^\infty w^{2r+1} g(w|D)dw = 0$ for r = 0, 1, 2, .... The even order moments are given by

$$\begin{aligned}\mu_{2r} &= 2\int_0^\infty w^{2r} e^{-2w} dw \\ &= \int_0^\infty (x/2)^{2r} e^{-x} dx = \Gamma(2r+1)/2^{2r} \qquad r = 1, 2, ... \\ &= (2r)!/2^{2r}\end{aligned} \qquad (A.4)$$

where $x = 2w$ and $\Gamma(\cdot)$ is the gamma function. For example, $\mu_2 = 1/2$, $\mu_4 = 3/2$, etc.

Note that $w = \sqrt{n}\,(\theta-\bar{y})/s\sqrt{2} = z/\sqrt{2}$, where $z = \sqrt{n}\,(\theta-\bar{y})/s$ is a standardized normal variable for the normal *conditional* posterior density for $\theta$ given $\sigma^2 = s^2$, $g_N(\theta|\sigma^2 = s^2,\bar{y}) \sim N(\bar{y}, s^2/n)$. For the conditional density, $E(z^2|\sigma^2 = s^2,D) = 1$, while for the unconditional density, $E(z^2|D) = 2E(w^2|D) = 1$ which, surprisingly, is equal to the conditional variance. However, $E(z^4|\sigma^2 = s^2,D) = 3$ in the conditional normal density for z while using the marginal density $Ez^4|D = 4Ew^4|D = 4\cdot 3/2 = 6$, two times the value of that for the conditional density. The fourth moment of z divided by the squared second moment, denoted by $\mu_4/\mu_2^2 = 6$ and the "excess" over that for the conditional normal density for z is 6 - 3 = 3. Thus the marginal double-exponential density for z is quite leptokurtic.

Note that $z = \sqrt{n}\,(\theta-\bar{y})/s$ has a univariate Student-t posterior density with $\nu$ = n-1 degrees of freedom if a standard normal likelihood function for $\theta$ and $\sigma$ and a diffuse prior $\pi(\theta,\sigma) \propto 1/\sigma$ were combined using Bayes' Theorem.[6] For this Student-t posterior density, we have: $Ez|D = 0$ for $\nu > 0$, $Ez^2|D = \nu/(\nu-2)$ for $\nu > 2$, and $Ez^4|D = 3\nu^2/(\nu-2)(\nu-4)$ for $\nu > 4$. Thus the excess for the Student-t based posterior density for z is: $\mu_4/\mu_2^2 - 3 = 6/(\nu-4)$, for $\nu > 4$, which for $\nu > 6$ is considerably less than 3, the excess of the double-exponential density for z. Thus, for n - 1 = $\nu > 6$ or n > 7, the double-exponential density is more leptokurtic. Also, as $\nu$ grows in value, the excess for the Student-t posterior density goes to zero whereas that for the double-exponential density has a constant value equal to 3.

Since $z = \sqrt{n}\,(\theta-\bar{y})/s = \sqrt{2}\,w$, the double-exponential posterior density for z is from (A.3).

$$p(z|D) = \left(1/\sqrt{2}\right)\exp\{-\sqrt{2}\,|z|\} \qquad -\infty < z < \infty \qquad (A.5)$$

---

[6]This well known result, probably first derived by Jeffreys, is presented in many works including Jeffreys (1988), Berger (1985), Box and Tiao (1973), Press (1989), and Zellner (1971).

Then for $c > 0$, $P_c \equiv \Pr(c < z < \infty|D) = \int_c^\infty f(z|D)dz = (1/2)\exp\{-\sqrt{2}\,c\}$ and thus $\ln 2P_c = -\sqrt{2}\,c$. For example for $P_c = .025$, $\ln .05 = -2.9957 = -1.4142c$ and $c = 2.118$. Thus $.025 = \Pr\{2.118 < z < \infty\}$ and from the symmetry of the density in (A.5), we have

$$\Pr\{-2.118 < z < 2.118|D\} = .95 \qquad (A.6)$$

This 95% interval for $z = \sqrt{n}\,(\theta-\bar{y})/s$, -2.118 to 2.118 implies that a 95% interval for $\theta$ is $\bar{y} \pm 2.118\,s/\sqrt{n}$. This interval is somewhat broader than a 95% interval based on the conditional normal posterior density for $\theta$ with $\sigma^2 = s^2$, namely $\bar{y} \pm 1.96\,s/\sqrt{n}$, the widths being $4.24\,s/\sqrt{n}$ in the former case and $3.92\,s/\sqrt{n}$ in the latter, about an 8% difference.